# Negative-index bi-anisotropic photonic metamaterial fabricated by direct laser writing and silver shadow evaporation

Michael S. Rill,[1,*] Christine Plet,[1] Michael Thiel,[1] Georg von Freymann,[1,2] Stefan Linden,[1,2] and Martin Wegener[1,2]

[1]*Institut für Angewandte Physik and DFG-Center for Functional Nanostructures (CFN), Universität Karlsruhe (TH), Wolfgang-Gaede-Str. 1, 76131 Karlsruhe, Germany*
[2]*Institut für Nanotechnologie, Forschungszentrum Karlsruhe in der Helmholtz-Gemeinschaft, 76021 Karlsruhe, Germany*

[*]*Corresponding author: michael.rill@physik.uni-karlsruhe.de*

We present the blueprint for a novel negative-index metamaterial. This structure is fabricated via three-dimensional two-photon direct laser writing and silver shadow evaporation. The comparison of measured linear optical spectra with theory shows good agreement and reveals a negative real part of the refractive index at around 3.85 μm wavelength - despite the fact that the metamaterial structure is bi-anisotropic due to the lack of inversion symmetry along its surface normal.

Metamaterials are tailored artificial materials composed of sub-wavelength metallic building blocks ("photonic atoms") that are densely packed into an effective material [1]. In this fashion, for example, an effective negative index of refraction ("optical antimatter") can be achieved [2,3]. Most metamaterials operating at optical frequencies have been fabricated using planar lithography techniques (e.g., electron-beam lithography or microcontact printing) and, hence, the resulting structures have essentially been planar – with few notable exceptions [4,5]. Recently, early work [6] has employed three-dimensional (3D) direct laser writing (DLW) and a combination of $SiO_2$ atomic-layer deposition (ALD) and silver chemical vapor deposition (CVD). Approaches like that appear to be very promising for the rapid prototyping of truly 3D metamaterials, however, theoretical blueprints for meaningful metamaterial structures compatible with these approaches still need to be developed. In this Letter, we propose (and then realize) a non-planar structure that can be made by DLW and silver shadow evaporation.

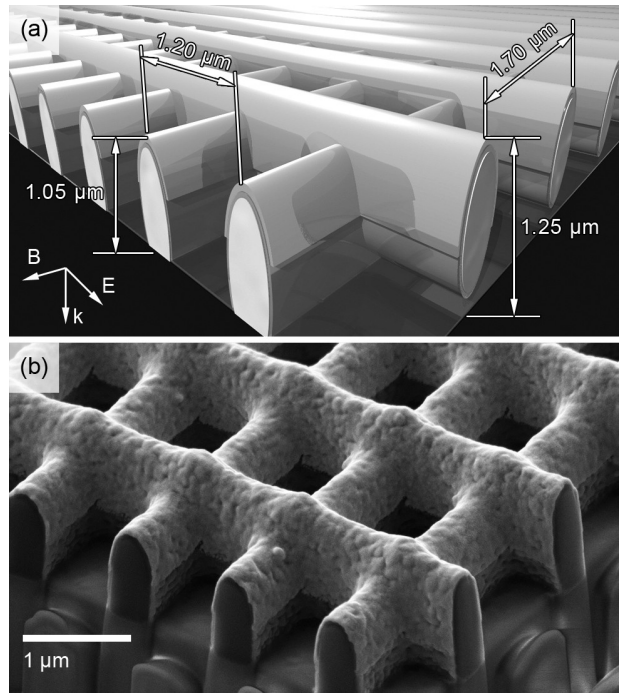

**Fig.1:** (a) Metamaterial design. The white regions are the polymer (SU-8) located on a glass substrate. The side walls of the polymer (encapsulated by $SiO_2$ via ALD) are coated with silver. The polarization of the incident electromagnetic field is illustrated on the lower left-hand side corner. (b) Oblique-view electron micrograph of a structure fabricated by direct laser writing and silver shadow evaporation that has been cut by a focused-ion beam (FIB) in order to reveal its interior. The complicated features visible underneath the glass-substrate surface are due to the FIB cutting and, hence, not relevant.

Sample design is illustrated in Fig.1 (a). First, a polymer (SU-8) template is made by DLW using a commercially available turn-key 3D DLW lithography system (Nanoscribe GmbH, see www.nanoscribe.de). Next, the template is coated with a thin layer of $SiO_2$ using an ALD process and metallized by high-vacuum electron-beam evaporation of silver. The surface normal and the axis of evaporation include a fixed angle of 65°. During evaporation, the azimuth angle is brought to four different positions such that the SU-8 side walls get coated as well (but, e.g., the substrate remains uncoated). This procedure leads to the final metal-dielectric composite structure shown in Fig.1 (b). The upside down "U" parts can be viewed as split-ring resonators [1], which deliver the required magnetic-dipole response. The intentionally elevated and elongated metal parts parallel to the incident electric-field vector deliver the negative electric permittivity.

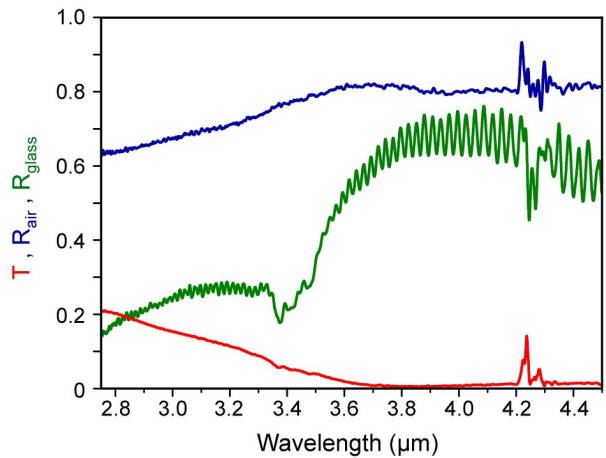

**Fig.2:** Measured linear-optical normal-incidence transmittance (red curve) and reflectance spectra of a structure similar to the one shown in Fig.1 (b). Note that the reflectance taken from the air side $R_{air}$ (blue curve) and that taken from the glass-substrate side $R_{glass}$ (green curve) are distinctly different.

Measured transmittance and reflectance spectra are shown in Fig.2. The transmittance spectrum exhibits a minimum at around 3.85 μm wavelength. The peaks in the spectra at around 4.2 μm wavelength are due to $CO_2$ absorption lines in the spectrometer and, hence, an artifact of the measurement. Notably, the normal-incidence reflectance spectrum taken from the air side is substantially different from that taken from the glass-substrate side, whereas the transmittance is the same for both sides within experimental uncertainty. This aspect is an immediate consequence of the fact that the overall structure (see Fig.1) lacks inversion symmetry along the optical axis (surface normal). (The rapid oscillations are due to Fabry-Perot interferences in the 170 µm thick glass substrate.) In order to further elucidate the physics underlying these measured optical spectra, we compare with theory.

Our corresponding numerical calculations for the actual nanostructure (for geometric dimensions, e.g., the lattice constants of 1.2 µm and 1.7 µm, respectively, see Fig.1 (a)) use a commercial finite-difference time-domain software (CST Microwave Studio) and literature parameters for the permittivity of SU-8, $\varepsilon_{SU-8} = 2.46$, that of the glass substrate, $\varepsilon_{SiO2} = 2.31$, and that of the silver, which is described by the usual free-electron Drude model with plasma frequency $\omega_{pl} = 1.37 \cdot 10^{16}$ s$^{-1}$ and collision frequency $\omega_{col} = 8.50 \cdot 10^{13}$ s$^{-1}$. The SiO$_2$ film thickness is 35 nm. The silver film thickness is 32 nm on all rod sides and 51 nm on top of the rods.

Calculated spectra that can be compared directly with experiment (Fig.2) are depicted in Fig.3 (a). The overall agreement with experiment is good. In particular, theory also shows a minimum of transmittance at around 3.85 µm wavelength. Remaining discrepancies between experiment and theory are very likely due to slight imperfections in sample fabrication (see Fig.1 (b)). Importantly, theory reproduces that the normal-incidence reflectance spectra taken from the air and the glass-substrate side, respectively, are quantitatively and qualitatively different.

As a result – as pointed out previously [6] on the basis of a bulk of literature on bi-anisotropy [7-9] – a description in terms of electric permittivity and magnetic permeability alone is generally not possible. Under our conditions, the relevant components of the electromagnetic fields are connected via

$$\begin{pmatrix} D \\ B \end{pmatrix} = \begin{pmatrix} \epsilon_0 \epsilon & -ic_0^{-1}\xi \\ +ic_0^{-1}\xi & \mu_0\mu \end{pmatrix} \begin{pmatrix} E \\ H \end{pmatrix} \qquad (1)$$

where $\varepsilon_0$ is the vacuum permittivity, $\mu_0$ the vacuum permeability, and $c_0$ the vacuum speed of light.

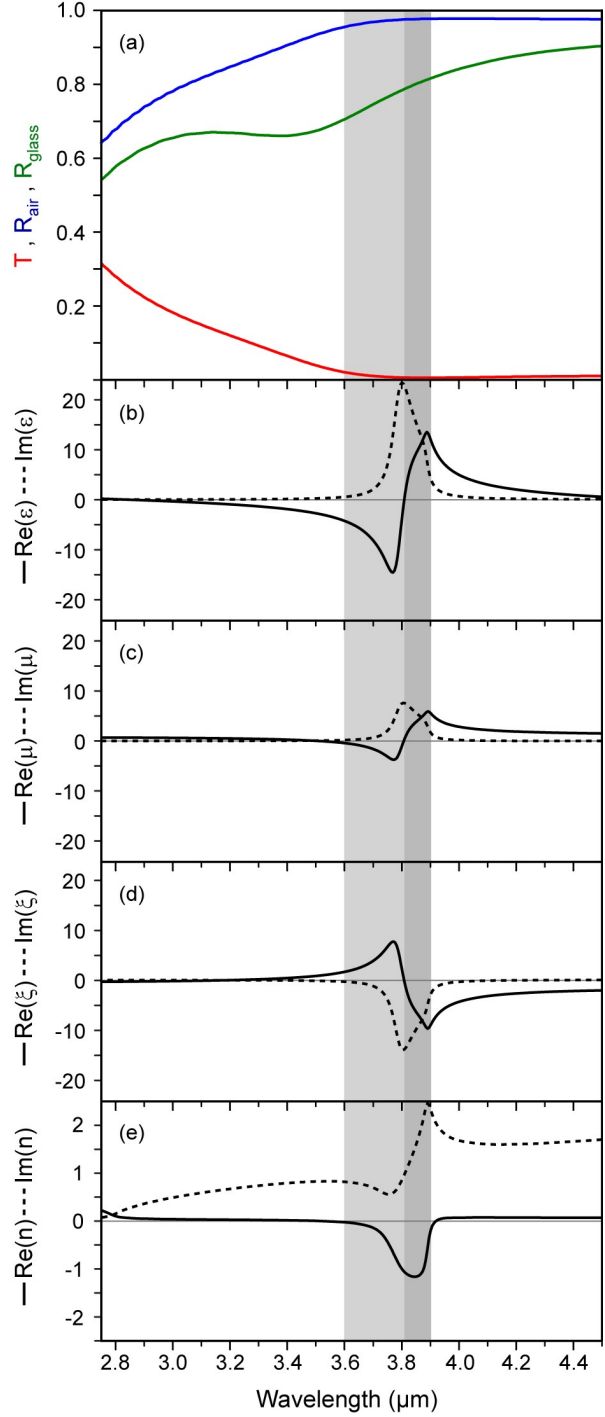

Fig.3: Calculated optical response of the structure shown in Fig.1 (a) for normal incidence of light. (a) Transmittance and reflectance spectra that can be compared directly with experiment (Fig.2). (b) Retrieved electric permittivity $\varepsilon$, (c) magnetic permeability $\mu$, (d) bi-anisotropy parameter $\xi$, and (e) refractive index $n$. The corresponding real parts are shown as solid curves, the imaginary parts as dashed curves. The gray backgrounds aim at clarifying the origin of the negative real part of $n$.

The electric permittivity $\varepsilon$ describes excitation of electric dipoles via the electric component of the incident light field, the magnetic permeability $\mu$ the excitation of magnetic dipoles via the magnetic component, and the bi-anisotropy parameter $\xi$ the excitation of magnetic dipoles via the electric component of the light field and vice versa. Due to reciprocity, the two corresponding off-diagonal terms of the above 2 × 2 matrix are strictly identical up to the minus sign. The complex refractive index $n$ is given by

$$n^2 = \epsilon\mu - \xi^2 \qquad (2)$$

The effective metamaterial parameters are retrieved by exactly matching the complex reflectance – which depends on from which side of the sample it is taken – and the complex transmittance (see Fig.3 (a)) – which does not depend on from which side it is taken due to reciprocity – of a bi-anisotropic slab of thickness 1.25 µm (see Fig.1 (a)) on a glass substrate to the corresponding quantities obtained from the numerical calculations for the actual nanostructure. The formulas required for this bi-anisotropic retrieval are explicitly given in [6].

Results of this retrieval are depicted in Figs.3 (b)-(e). A negative real part of the refractive index $n$ is achieved from around 3.6 µm to 3.9 µm wavelength. Notably, the negative real part of $n$ is partly connected to the real parts of electric permittivity and magnetic permeability being negative (see light gray area) and partly due to the real part of the bi-anisotropy parameter being negative (see dark gray area). The corresponding imaginary part of n can be translated into a maximum figure of merit of FOM = Re($n$) / Im($n$) = 1.3, which is comparable to double-fishnet-type negative-index photonic metamaterials made via electron-beam lithography [2,3]. In this spectral range, a description in terms of an effective medium is justified because the lattice constant is smaller than half the wavelength.

In summary, we have designed, fabricated, and characterized a novel negative-index metamaterial that is bi-anisotropic. The design is robust against fabrication tolerances and imperfections. While our particular fabrication approach takes advantage of silver shadow evaporation, hence cannot directly be extended to an arbitrary number of functional layers into the third dimension, the presented results do provide us with significant further encouragement that state-of-the-art 3D direct laser writing can lead to very high quality photonic (negative-index) metamaterials operating at near-infrared wavelengths. Also, this example has shown that a negative refractive index can be achieved in a bi-anisotropic structure, whereas our previous results on a different structure [6] may have suggested the opposite.

## Acknowledgments

We thank Costas M. Soukoulis for stimulating discussions. We acknowledge financial support provided by the Deutsche Forschungsgemeinschaft (DFG) and the State of Baden-Württemberg through the DFG-Center for Functional Nanostructures (CFN) within subprojects A1.4 and A1.5. The project PHOME acknowledges the financial support of the Future and Emerging Technologies (FET) programme within the Seventh Framework Programme for Research of the European Commission, under FET-Open grant number 213390. The METAMAT project is supported by the Bundesministerium für Bildung und Forschung (BMBF). The research of S.L. is further supported through a "Helmholtz-Hochschul-Nachwuchsgruppe" (VH-NG-232), the research of G.v.F through the DFG Emmy-Noether fellowship (FR 1671/4-3), the PhD education of M.S.R. and M.T. by the Karlsruhe School of Optics & Photonics (KSOP).